\documentclass[conference]{IEEEtran}
\IEEEoverridecommandlockouts

\usepackage{cite}
\usepackage{amsmath,amssymb,amsfonts}
\usepackage{algorithm}
\usepackage{algpseudocode}
\usepackage{graphicx}
\usepackage{textcomp}
\usepackage{xcolor}
\usepackage{float}  
\usepackage{hyperref}
\hypersetup{hidelinks}
\usepackage{mathptmx}
\usepackage{booktabs}     
\usepackage{siunitx}      
\usepackage{array}        
\usepackage{caption}      
\usepackage{subcaption}   

\newcolumntype{P}[1]{>{\centering\arraybackslash}p{#1}}
\newcolumntype{M}[1]{>{\centering\arraybackslash}m{#1}}

\usepackage{numprint}

\usepackage{placeins} 

\def\BibTeX{{\rm B\kern-.05em{\sc i\kern-.025em b}\kern-.08em
    T\kern-.1667em\lower.7ex\hbox{E}\kern-.125emX}}
\begin{document}

\title{QoS-aware Scheduling of Periodic Real-time Task Graphs on Heterogeneous Pre-occupied MECs\\
}

\author{\IEEEauthorblockN{1\textsuperscript{st} Ashutosh Shankar}
\IEEEauthorblockA{\textit{Department of Chemical Engineering
} \\
\textit{Indian Institute Of Technology Kharagpur}\\
Kharagpur, India\\
 \href{mailto:ashutoshshankar74@gmail.com}{ashutoshshankar74@gmail.com}
}
\and
\IEEEauthorblockN{2\textsuperscript{nd} Astha Kuamari}
\IEEEauthorblockA{\textit{Department of Chemical Engineering
} \\
\textit{Indian Institute Of Technology Kharagpur}\\
Kharagpur,India \\
\href{mailto:astha.kumari1103k@gmail.com}{astha.kumari1103k@gmail.com}
}

}

\maketitle

\begin{abstract}
In latency-sensitive applications, efficient task scheduling is crucial for maintaining Quality of Service (QoS) while meeting strict timing constraints. This paper addresses the challenge of scheduling periodic tasks structured as directed acyclic graphs (DAGs) within heterogeneous, pre-occupied Mobile Edge Computing (MEC) networks. We propose a modified version of the Heterogeneous Earliest Finish Time (HEFT) algorithm designed to exploit residual processing capacity in pre-occupied MEC environments. Our approach dynamically identifies idle intervals on processors to create a feasible hyperperiodic schedule that specifies an allocated virtual machine (VM), task version, and start time for each task. This scheduling strategy maximizes the aggregate QoS by optimizing task execution without disrupting the existing periodic workload, while also adhering to periodicity, precedence, and resource constraints. Experimental results demonstrate that our method achieves enhanced load balancing and resource utilization, highlighting its potential to improve performance in heterogeneous MEC infrastructures supporting real-time, periodic applications.

\end{abstract}
\vspace{0.2cm}

\section{Introduction}
With the growth of cloud computing, resource heterogeneity has become critical to improving the performance and efficiency of cloud-based systems. Resource heterogeneity[9] is the integration of several specialized processing units inside a cloud architecture, each designed to handle certain types of workloads efficiently. These resources include central processing units (CPUs) for general-purpose computing, graphics processing units (GPUs) for parallel processing tasks, tensor processing units (TPUs) for deep learning, and field-programmable gate arrays (FPGAs) for customizable, low-latency applications. By leveraging this diversity, cloud platforms enable diverse applications in the domains of artificial intelligence (AI), scientific simulations, and big data analytics to operate at higher efficiencies, optimized for performance, cost, and adaptability. Unlike traditional homogeneous systems, where all processing units are identical [1] and perform at similar levels, heterogeneous systems enable cloud users to match specific workload requirements to the most suitable resources, improving overall performance and cost efficiency.

In heterogeneous cloud environments, the challenge is to efficiently schedule and allocate tasks across diverse resources to achieve optimal performance, reduced idle time, and improved cost efficiency, while maximizing the quality of service (QoS) delivered by the system. This paper addresses this problem by focusing on a static scheduling approach, which is particularly suitable for predictable and periodic workloads. Static scheduling allows task assignments to be determined prior to execution, providing a high degree of predictability and control over resource distribution based on prior knowledge of task requirements and system capacity.

Unlike dynamic scheduling, which allocates tasks at runtime and adapts to changing system states, static scheduling is ideal for applications where task patterns are known in advance, allowing for optimized resource allocation and quality maximization in a cloud environment with pre-occupied processors. Our approach specifically aims to maximize the quality by ensuring tasks are allocated to the most suitable resources, considering both performance and service-level objectives.

Directed Acyclic Graphs (DAGs) are widely used in task scheduling for cloud computing as they represent workflows where nodes denote tasks and edges represent dependencies between them. In a DAG, each task must follow its precedence constraints, ensuring that a task does not start until its dependencies are fulfilled. This structure is particularly useful for modeling workflows in heterogeneous environments, as it enables visualization and tracking of tasks with dependencies, which is crucial for efficient scheduling. By representing workflows as a DAG, we can better manage the complexity of resource allocation and optimize task execution order, leading to improved performance, minimized idle time, and enhanced quality of service.

One popular scheduling algorithm for heterogeneous computing environments, especially when dealing with DAGs, is the Heterogeneous Earliest Finish Time (HEFT) algorithm. HEFT is a static scheduling algorithm designed to achieve optimal task allocation by minimizing the makespan — the total execution time for all tasks in the workflow. HEFT prioritizes tasks based on their level, which is a measure of task importance derived from its position in the DAG and the weight of the edges leading to it. Tasks with higher levels (often near the top of the DAG) are prioritized for assignment. For each prioritized task, HEFT selects the processor that minimizes the task's earliest possible finish time while considering both task dependencies and processor heterogeneity. This approach is especially suitable for heterogeneous environments, as it efficiently leverages diverse processing resources to meet the workflow's scheduling constraints, thereby maximizing the overall quality of the service provided.

This paper proposes a new scheduling algorithm that builds upon and modifies the HEFT algorithm to better suit the unique demands of heterogeneous cloud environments. This modified algorithm aims to improve task scheduling by minimizing the makepan and enhancing the effective use of diverse resources, specifically maximizing quality and ensuring that service-level objectives are met. The effectiveness of our proposed algorithm is demonstrated through a practical example.
The remaining parts of this paper are organized as follows: Section II presents the Literature review. In Section III, we describe the task scheduling problem and its formulation. Section IV explains the problem statement. In Section V, we introduce our proposed scheduling algorithm, which is a modification of the HEFT algorithm, and demonstrate its performance through an example. Section VI presents the results of our work. Finally, Section VII concludes the paper and discusses future research directions.

\section{Related Work}

\subsection{Maintaining the Integrity of the Specifications}

In this section, we conduct a survey of related work on task scheduling techniques in cloud computing, highlighting significant advancements that enhance scheduling efficiency and Quality of Service (QoS) in heterogeneous environments. Task scheduling is a critical component in managing the extensive virtualized resources available in cloud computing, where manual scheduling is impractical due to the scale of operations. The paper [2] provides a comprehensive overview of various scheduling strategies and metrics suitable for cloud environments. It categorizes existing research based on methods, applications, and parameter-based measures while identifying limitations in current methodologies. This survey can be used to highlight precisely where new algorithms are needed for an enhanced cloud scheduling performance. Varied techniques including heuristics[7], mathematical operations, and machine learning models as well as evolutionary algorithms – such as the genetic algorithm, [8]ant colony optimization algorithm, cuckoo search algorithm, and swarm intelligence algorithm – can be used to develop hybrid algorithms that can perform more effectively compared to existing approaches. 

The study [3] proposes an innovative two-stage strategy aimed at enhancing task scheduling performance and minimizing unreasonable task allocation in cloud environments. In the first stage, a job classifier inspired by Bayes' design principle classifies tasks based on historical scheduling data, allowing for the pre-creation of virtual machines (VMs) of different types. This approach reduces delays associated with VM creation during task execution. In the second stage, tasks are dynamically matched with suitable VMs, effectively improving load balancing and overall scheduling performance compared to traditional methods. The proposed framework aims to minimize task execution costs while meeting deadlines by efficiently utilizing VMs and reducing waiting times.

One notable approach focuses on scheduling user tasks in cloud environments while minimizing energy consumption and adhering to QoS requirements. This study introduces a model that analyzes the state of running tasks using a QoS prediction method based on an ARIMA model optimized with a Kalman filter. The authors propose a hybrid scheduling policy that combines Particle Swarm Optimization (PSO) and Gravitational Search Algorithm (GSA) to optimize task scheduling based on QoS status analysis. Experimental results indicate that the proposed HPSO algorithm achieves a 16.51\% reduction in resource consumption compared to the original hybrid algorithm, with minimal service-level agreement (SLA) violations. This[6] highlights the importance of integrating QoS considerations into task scheduling strategies for improved resource efficiency, addressing a critical gap found in many existing approaches. 

Additionally, performance-effective and low-complexity task scheduling algorithms provide insights into efficient scheduling solutions[4]. The paper presents two novel algorithms: the Heterogeneous Earliest-Finish-Time (HEFT) algorithm and the Critical-Path-on-a-Processor (CPOP) algorithm. These algorithms prioritize tasks based on their rank values to minimize execution time while maintaining low scheduling costs.

A common drawback of previous approaches is their varying degrees of consideration for load balancing and QoS requirements. The Enhanced HEFT (E-HEFT) algorithm, as presented in [5] addresses some of these deficits by achieving a balanced load across machines and minimizing the makespan of workflow applications under specified budget constraints. This paper proposes a modified HEFT algorithm tailored for latency-sensitive, periodic applications represented as Directed Acyclic Graphs (DAGs). It dynamically identifies idle intervals in the preoccupied Mobile Edge Computing (MEC) network, leveraging the residual capacity to improve the aggregate QoS acquired by the application. Our approach generates a hyperperiodic schedule that determines, for each task, an allocated VM, a selected task version, and a suitable start time while ensuring that periodicity, precedence, and resource constraints are met. This enhancement improves load distribution and resource utilization within heterogeneous computing environments without disrupting the existing periodic workload.

\section{Problem Formulation}
\subsection{System Model}
The platform considers a geographically distributed set of heterogeneous MEC (Mobile Edge Computing) host servers, each represented as \( H_i \), capable of communicating over a fully interconnected network. These MEC hosts consist of virtual machines (VMs) denoted by \( V_i \) that can execute dynamically arriving periodic real-time tasks. The tasks are represented as a directed acyclic graph (DAG), where each node corresponds to a computation task, and edges represent data dependencies between tasks. We aim to ensure that tasks meet their timing constraints while maximizing Quality of Service (QoS) through efficient scheduling.

The set of all VMs that can execute a specific periodic task graph G across MEC hosts is represented by [ S = { \(V_1\), \(V_2\), ..., \(V_{|S|}\) ] In a fully connected network, VMs can communicate with each other through distinct links characterized by unique bandwidths ( \(b_{nm}\) ), where ( \(V_n\), \(V_m\) in S ). Figure 2 illustrates a sample platform with a 3-VM configuration, showing potential connections between VMs and corresponding link bandwidths.

\textbf{VM Pre-Occupation Model}
Our approach assumes that, for the entire duration up to the deadline of the periodic application \( G \) that needs to be scheduled, both processors in the system are already preoccupied with an existing scheduled workload. This gives a snapshot of the platform availability, showing all the free intervals in white and occupied intervals in grey at the arrival of \( G \).

The event queue \( EQ_p \) of processor \( P_p \) is represented as a list of free slots, with each slot represented as \([sp_j, dp_j]\), where \( sp_j \) is the start time, and \( dp_j \) is the duration of the free slot. The list is sorted in increasing order of start times \( sp_j \). Without loss of generality, the start time of the first slot in each processor's event queue is either the release time of the periodic application \( G \) to be scheduled or the start time of the earliest free slot subsequent to that release time.

The preoccupied scheduling event queues correspond to the hyper-periodically repeating workload of persistent periodic applications, ensuring that periodic constraints and dependencies are respected within the created hyperperiod.

\textbf{Operations on Event Queue}
To maximize the aggregate QoS of the application GGG, the schedule may be modified multiple times to increase the QoS of a task, which in turn requires tasks to be tentatively deallocated and reallocated to a different idle slot in a VM. After every deallocation and reallocation of tasks, the following operations on the EQ of the corresponding VM may be performed: i) merging two or more consecutive idle slots after deallocating a slot, and ii) splitting an idle slot, after partial allocation of the slot to a task, respectively. After every merge or split operation, resulting idle slots are added to the EQ, while still maintaining the original order of increasing start times of idle slots. These operations can be performed in linear time.

\textbf{Background}
Traditional scheduling methods focus on allocating tasks to processors to maximize resource utilization and minimize execution time. However, integrating new tasks with those already running on preoccupied processors presents unique challenges. In these cases, it is essential to address both the frequency of tasks and the need to optimize quality without altering established processor mappings. This is particularly relevant in real-time applications such as traffic monitoring, where new tasks—such as incident response or signal adjustments—must coexist with ongoing processes to ensure continuous, real-time functionality.
In a traffic monitoring system, edge servers are typically positioned at key locations within a network, responsible for running periodic tasks such as vehicle counting, speed analysis, and anomaly detection. These tasks demand high-quality scheduling to avoid delays in capturing critical data or responding to events. When sudden events, like accidents or congestion, trigger additional processing tasks, these must be integrated into the existing task schedules without affecting the ongoing monitoring tasks. Any disruption in processor mappings could reduce response efficiency, affecting overall system quality and reliability. In such environments, scheduling solutions that allow seamless integration of new tasks are crucial to maintaining high service standards.

Among the scheduling algorithms for heterogeneous systems, the Heterogeneous Earliest Finish Time (HEFT) algorithm has been one of the most frequently cited and used because of its simplicity and good performance. HEFT is a natural extension of the classical list scheduling algorithm for homogeneous systems to cope with heterogeneity. It outperforms other comparable algorithms in terms of minimization of the execution time.HEFT’s ability to prioritize and sequence tasks across multiple processors based on the earliest completion time makes it suitable for balancing task quality and execution constraints, making it an ideal choice for quality-focused, multi-task scheduling on shared resources.

\begin{figure}
    \centering
    \includegraphics[width= 5cm, height= 7 cm]{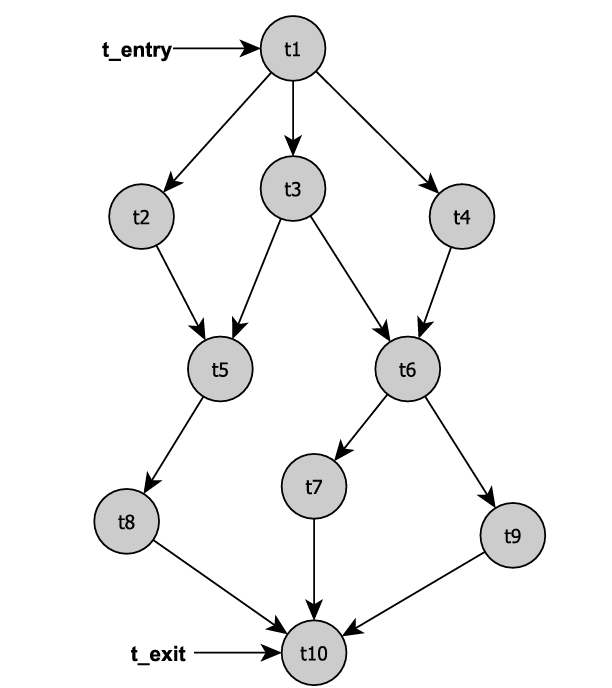}
    \caption{An example of workflow application}
    
\end{figure}

\begin{figure}
    \centering
    \includegraphics[width= 5cm, height= 6 cm]{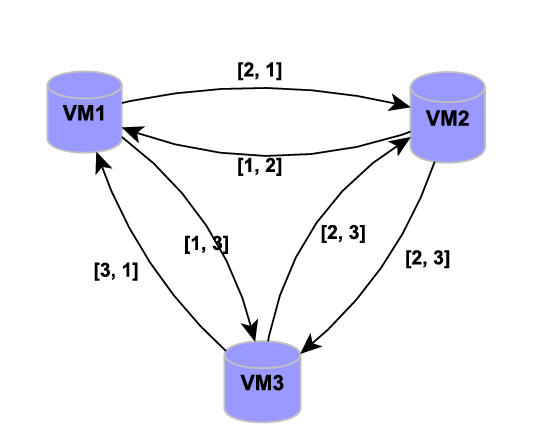}
    \caption{Virtual machines}
    
\end{figure}
\subsection{Application model}
In a periodic directed acyclic graph (DAG), tasks are arranged in a recurring sequence that repeats after each specified period. Let \( T = \{ t_1, t_2, t_3, \dots, t_n \} \) represent the set of tasks, and let \( E \) denote the set of directed edges that define dependencies between tasks within each period. Each edge \( e : (t_i, t_j) \) in graph \( G \) signifies a precedence constraint, where \( t_i \) is the parent of \( t_j \) and \( t_j \) cannot commence until \( t_i \) has completed. If there is data transfer from \( t_i \) to \( t_j \), then \( t_j \) can only start once all required data from \( t_i \) has been received.

Within each period, tasks without any preceding parent tasks are referred to as entry tasks, denoted as \( t_{\text{entry}} \), while those with no succeeding child tasks are termed exit tasks, denoted as \( t_{\text{exit}} \). At the end of each period, the workflow repeats, initiating from the entry tasks again.
A sample workflow is shown in figure 2.
\vspace{0.2cm}
\section{Problem Statement}
Given a latency-sensitive periodic application structured as a DAG G = (T, E), having a period D, determine a feasible schedule which specifies for each task: i) an allocated VM, ii) a selected task version, and iii) a suitable start time. The scheduling objective is to obtain a hyperperiodic schedule in order to maximize aggregate QoS acquired by G with the following constraints: 

i. It does not jeopardise the existing periodic workload while utilizing the residual capacity of a preoccupied MEC network in order to maximize aggregate QoS acquired  by G, 

ii. The schedule meets periodicity, precedence, and resource constraints.

\textbf{An Example}

To analyze the scheduling of periodic tasks with enhanced quality adjustments, we first outline the configuration of a preoccupied processor with recurring tasks, a Directed Acyclic Graph (DAG) of tasks to be scheduled within the gaps left by the processor's pre-scheduled tasks, and then proceed through detailed scheduling steps that include enhancements in task quality.

In the example provided, preoccupied processor has an initial period of 8. A new task DAG (Directed Acyclic Graph) with a period of 12 is introduced, requiring scheduling across processors \(V1\) and \(V2\).
\begin{figure}
    \centering
    \includegraphics[width= 6cm, height= 8cm]
    {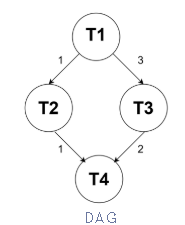}
        \caption{Task DAG}
    
\end{figure}
\begin{table}[h!]
\centering
\renewcommand{\arraystretch}{1} 
\setlength{\tabcolsep}{10pt} 
\begin{tabular}{|c|c|c|c|c|}
\hline
      & \multicolumn{2}{c|}{V1} & \multicolumn{2}{c|}{V2} \\ \hline
      & Q1 & Q2 & Q1 & Q2 \\ \hline
T1    & 1  & 2  & 1  & 2  \\ \hline
T2    & 2  & 3  & 2  & 3  \\ \hline
T3    & 2  & 3  & 2  & 3  \\ \hline
T4    & 1  & 2  & 1  & 2  \\ \hline
\end{tabular}
\vspace{0.2cm}
\caption{Tasks and their quality levels}
\label{tab:example_table}
\end{table}

\textbf{Initial Setup}
\begin{itemize}
    \item The \textbf{preoccupied processor} has periodic tasks with a period of \(8\), representing existing workloads.
    \item The new \textbf{task graph}, represented by \(T1\), \(T2\), \(T3\) and \(T4\), is scheduled on processors \(V1\) and \(V2\). Each task is associated with specific quality parameters (\(Q1\) and \(Q2\)) on both processors.
\end{itemize}

\textbf{Hyperperiod Calculation:}
The \textbf{hyperperiod} is calculated as the least common multiple (LCM) of the existing period (8) and the new period (12). This ensures that task schedules for both periods align seamlessly over time.

\[
\text{Hyperperiod} = \text{LCM}(8, 12) = 24
\]
\begin{figure}[H]
    \centering
    \includegraphics[width=0.45\textwidth]{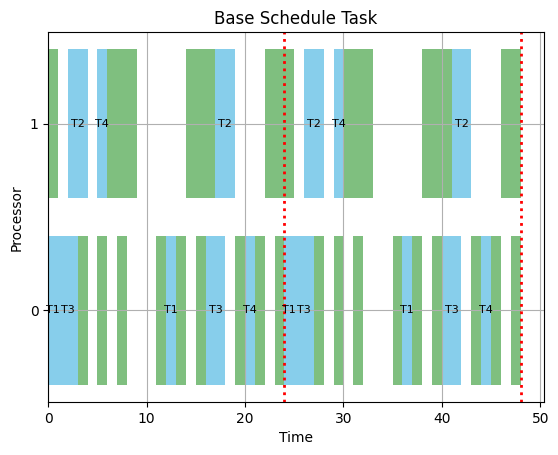}
    \caption{Base quality}
    \label{fig:base_quality}
\end{figure}


\begin{figure}[H]
    \centering
    \includegraphics[width=0.45\textwidth]{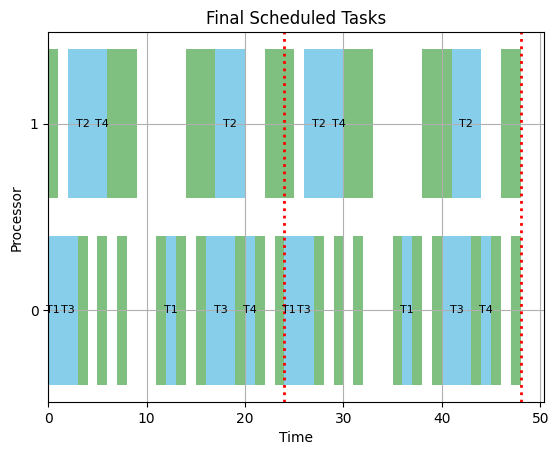}
    \caption{Enhanced quality}
    \label{fig:enhanced_quality}
\end{figure}

\section{Methodology}

This section introduces a novel scheduling algorithm for optimizing task execution quality and efficiency in heterogeneous processors handling periodic Directed Acyclic Graphs (DAGs). The framework comprises two core components: a Modified HEFT algorithm for scheduling DAGs on preoccupied processors while maintaining periodicity, and a Periodic Scheduler that ensures tasks across DAGs do not overlap and repeat seamlessly after each hyperperiod. By integrating DAGs into processors with preallocated tasks, the algorithm strategically schedules new tasks into idle time gaps without disrupting existing task sequences, initially assigning base quality levels. Subsequently, it maximizes execution quality by dynamically adjusting task quality levels to utilize idle slots optimally, thereby enhancing resource utilization and output quality without extending the schedule’s makespan. This dual-phase approach balances computational efficiency with quality improvements, ensuring robust periodic task management in heterogeneous environments.

\subsection{Modified HEFT: Scheduling Secondary DAG Tasks on the preoccupied processors in available GAPs}

The Modified HEFT algorithm extends the traditional Heterogeneous Earliest Finish Time (HEFT) approach to schedule secondary Directed Acyclic Graph (DAG) tasks on preoccupied processors by leveraging available computational gaps within predefined cycles. Like standard HEFT, it begins by calculating task ranks using upward rank values (average computation cost plus communication delays) and sorts tasks in descending priority order. However, unlike the original algorithm, it incorporates constraints such as processor preoccupation, cycle boundaries (CycleStartTime and CycleEndTime), and real-time resource availability. This adaptation allows it to operate within dynamic environments where processors already have allocated tasks with intermittent idle periods.

\subsubsection*{Step 1: Upward Rank Calculation and Task Sorting}

The algorithm begins by computing the \textit{upward rank} for each task, a critical path metric derived from the average computation cost across all processors and the communication delays between dependent tasks. For a task $n_i$, the upward rank is defined as:

\begin{equation}
\text{rank}(n_i) = \overline{w_i} + \max_{n_j \in \text{succ}(n_i)} \left( \overline{c_{i,j}} + \text{rank}(n_j) \right)
\label{eq:upward-rank}
\end{equation}

where $\overline{w_i}$ is the average computation cost of task $n_i$, and $\overline{c_{i,j}}$ is the average communication cost between task $n_i$ and its successor $n_j$. 

Tasks are then sorted in descending order of their rank values, prioritizing those that have a higher impact on overall workflow completion.

\subsubsection*{Step 2: Gap-Aware Earliest Start Time (EST) Computation}

During processor assignment, the algorithm evaluates the \textit{Earliest Start Time (EST)} by balancing the following constraints:

\begin{itemize}
    \item \textbf{Cycle boundaries:} Tasks must start after \texttt{CycleStartTime} and finish before \texttt{CycleEndTime}.
    
    \item \textbf{Predecessor dependencies:} EST accounts for the latest finish time of all predecessors, including inter-task communication delays.
    
    \item \textbf{Idle gaps:} On preoccupied processors, EST is dynamically adjusted to fit tasks into available intervals between existing allocations.
\end{itemize}

For example, if a processor has a gap between two scheduled tasks, the algorithm computes EST as:

\begin{align}
\text{EST} = \max( & \text{CycleStartTime},( \text{CommunicationDelay}+ \nonumber \\
                       & \text{LatestPredecessorFinishTime} ))
\end{align}

The corresponding \textit{Earliest Finish Time (EFT)} is then calculated as:
\[
\text{EFT} = \text{EST} + \text{task.ComputationCost}
\]
Only gaps where $\text{EFT} \leq \text{CycleEndTime}$ are considered valid for scheduling.


\begin{algorithm}[H]
\caption{MODIFIED HEFT}
\begin{algorithmic}[1]

\State \textbf{Input:}
\State \hspace{1em} \texttt{ DAG, PreOccupiedProcessorsWithGaps, DAGPeriod, CycleStartTime, CycleEndTime}

\State \textbf{Output:}
\State \hspace{1em} \texttt{Scheduled DAG with enhanced quality on given processors}

\State \textbf{// Step 1: Initial Scheduling with Base Quality}
\State Compute the rank value of each task using the HEFT ranking method
\State $SortedTasks \gets$ Tasks sorted in decreasing order of rank values

\For{each $task$ in $SortedTasks$}
    \State $bestFinishTime \gets \infty$
    \State $selectedProcessor \gets None$
    
    \For{each $processor$ in $PreOccupiedProcessorsWithGaps$}
        \State $EST \gets$ max($CycleStartTime$, $LatestPredecessorFinishTime + CommunicationDelay$)
        \State $EFT \gets EFT + task.ComputationCost$
        
        \If{$EFT < CycleEndTime$ \textbf{and} $EFT < bestFinishTime$}
            \State $bestFinishTime \gets EFT$
            \State $selectedProcessor \gets processor$
        \EndIf
    \EndFor
    
    \If{$selectedProcessor \neq None$}
        \State Assign $task$ to $selectedProcessor$
        \State Update $selectedProcessor$ schedule
    \Else
        \State \textbf{Print} "Schedule Generation Failure"
        \State \Return
    \EndIf
\EndFor

\State \textbf{// Step 2: Enhance Quality of Schedule}
\For{each $task$ in DAG}
    \If{$task.processor == None$}
        \State \textbf{Print} "Task", $task.id$, "not scheduled."
        \State \Return
    \EndIf

    \State Remove $task$ from $task.processor$ schedule

    \State Try scheduling with higher duration (enhanced quality level)

    \State $newStartTime \gets$ compute earliest possible start time considering:
    \State \hskip1em base schedule start time and available processor gaps
    \State $newFinishTime \gets newStartTime + task.EnhancedComputationCost$

    \If{$newFinishTime < CycleEndTime$}
        \State Update $task$ duration and reschedule on the processor
        \State Update $processor$ schedule accordingly
    \Else
        \State Revert $task$ to original base schedule
    \EndIf
\EndFor

\end{algorithmic}
\end{algorithm}


\subsubsection*{Step 3: Dependency-Aware Scheduling of Secondary Tasks}

Tasks in the secondary DAG are scheduled in an order that respects their dependency hierarchy. For each task in the secondary DAG, the Earliest Start Time (EST) is calculated based on the identified gaps and the completion of any predecessor tasks. Each task is scheduled within the first available gap that accommodates its computation time and does not overlap with pre-existing tasks.

The algorithm assesses each task in the secondary DAG and attempts to place it in the smallest gap that satisfies the dependency constraints. This process ensures that tasks execute in the correct order without affecting the initial schedule. By filling idle processor times with secondary tasks, this phase maximizes the effective utilization of processing resources without altering the exiting timeline.

\subsubsection*{Step 4: Quality Levels and Time Requirements}

Each task is associated with a set of quality levels, where each successive level requires more processing time.
\begin{itemize}
    \item \textbf{Baseline quality:} Default execution time.
    
    \item \textbf{Enhanced quality:} Execution time increases.
\end{itemize}

These levels allow the algorithm to adjust a task’s quality based on available time without requiring a complete rescheduling.

\subsubsection*{Step 5: Quality Enhancement}

In the next part, algorithm attempts to upgrade the quality level of each scheduled task. It examines each task’s scheduled slot and checks for idle gaps immediately following the task. If there is enough time within this gap to upgrade the task
to a higher quality level, the task is rescheduled with this increased quality. This
process is repeated for each task, maximizing quality in a stepwise manner while
adhering to time constraints.

For every task, the algorithm removes the task from its base schedule and tries to re-schedule it with a higher computation cost (corresponding to enhanced quality).

A new start time is computed considering the original schedule and the available gaps on the assigned processor.

If the task can finish before the cycle end time with the enhanced duration, it is updated and reinserted into the processor's schedule.

If not, the task is reverted back to its original configuration to maintain feasibility.

The iterative nature of this phase ensures that the quality of each task is optimized individually, without causing overlaps or violating dependency constraints. By strategically utilizing remaining idle gaps for quality enhancements, this phase significantly improves the output quality without increasing the total execution time of the schedule.

\subsection{ Periodic Scheduler}

The Periodic Scheduler manages scheduling periodic DAGs on a preoccupied processor by harmonizing their individual cycles into a shared hyperperiod-the least common multiple (LCM) of both the DAGs. This ensures the combined schedule repeats predictably without conflicts.

\subsubsection*{Step 1: Hyperperiod-Driven Scheduling}

When a new DAG (period $P_\text{new}$) is added to a processor already running a DAG with period $P_\text{existing}$, the scheduler calculates the hyperperiod as:

\[
\text{HyperPeriod} = \text{LCM}(P_\text{existing}, P_\text{new})
\]

This represents the smallest interval after which both DAGs’ execution patterns align. For example, if $P_\text{existing} = 3\,\text{ms}$ and $P_\text{new} = 4\,\text{ms}$, the hyperperiod is $12\,\text{ms}$. All task instances across both DAGs are scheduled within this window, and the pattern repeats indefinitely.

\begin{algorithm}[H]
\caption{Periodic Scheduler for Multiple DAGs with Enhanced Quality}
\begin{algorithmic}[1]
\State \textbf{Input:}
\State \hspace{1em} \texttt{ DAGs[], PreOccupiedProcessorsWithGaps}
\State \textbf{Output:}
\State \hspace{1em} \texttt{ New schedules of periodic DAGs on available processors}

\State $\textbf{HyperPeriod} \gets LCM($period of each DAG in DAGs$)$

\For{each $dagIndex$, $dag$ in DAGs}
    \State $dagPeriod \gets dag.Period$
    \State $NumberOfCycles \gets 2 \times \dfrac{HyperPeriod}{dagPeriod}$

    \For{$instance = 0$ to $NumberOfCycles - 1$}
        \State $CycleStartTime \gets \max(instance \times dagPeriod,\ LastEndTimes[dagIndex])$
        \State $CycleEndTime \gets CycleStartTime + dagPeriod$

        \State Create a copy of the DAG with unique task IDs
        \State Remap communication costs to match new task IDs

        \State \textbf{// Call Modified HEFT for base quality scheduling}
        \State \Call{\textbf{ModifiedHEFT}}{$CopiedDAG$, PreOccupiedProcessorsWithGaps, dagPeriod, CycleStartTime, CycleEndTime}
    \EndFor
\EndFor
\end{algorithmic}
\end{algorithm}

\subsubsection*{Step 2: Cycle-Based Instance Generation}

For each DAG, the scheduler generates the number of instances within the hyperperiod as:

\[
\text{NumberOfCycles} = \frac{\text{HyperPeriod}}{P_\text{DAG}}
\]

Each instance is confined to a cycle window ($\text{CycleStartTime}$ to $\text{CycleEndTime}$), ensuring temporal isolation. The scheduler avoids overlaps by enforcing rigid deadlines and prioritizing gap utilization via Modified HEFT.

\subsubsection*{Step 3: Task Placement in Preoccupied Processors}

{Gap Identification:} The processor’s existing tasks create idle intervals between executions.

{Modified HEFT Integration:} For each new DAG instance, tasks are placed into gaps that satisfy:
\begin{itemize}
    \item Dependencies (predecessor finish times + communication delays).
    \item Cycle boundaries (no deadline violations).
\end{itemize}

{Pattern Repetition:} After the hyperperiod, the entire schedule restarts, maintaining deterministic behavior

\section{Results and Experimental Evaluation}

This section presents an in-depth evaluation of the proposed scheduling methodology for latency-sensitive periodic DAG applications deployed over a resource-constrained Multi-access Edge Computing (MEC) environment. Each DAG $G = (T, E)$ comprises $n$ computational tasks (nodes), with directed edges denoting task precedence and communication dependencies. The scheduling objective is to maximize the aggregate Quality of Service (QoS), modeled as the \textbf{normalized reward}, under tight constraints of periodicity, resource availability, and inter-task dependencies.

\subsection{Normalized Reward Metric}

To quantify scheduling quality, we define the \textbf{Normalized Reward (NR)} as a percentage of the maximum achievable reward for a given DAG:
\begin{equation}
\text{NR} (\%) = \frac{R_{\text{ACT}}}{R_{\text{MAX}}} \times 100
\end{equation}
where:
\begin{itemize}
    \item $R_{\text{ACT}}$ is the actual cumulative reward obtained across all tasks in the DAG using the selected service levels and feasible resource assignments,
    \item $R_{\text{MAX}}$ is the theoretical maximum reward, achieved only when every task is executed using the highest service level (e.g., lowest latency, highest resolution).
\end{itemize}
The reward for each task is influenced by both its assigned service level and whether it meets its timing and precedence constraints. If any task violates its deadline or misses dependency enforcement, its reward is nullified, thus penalizing the overall $R_{\text{ACT}}$.

\subsection{Experimental Setup}
For each DAG size $n \in \{10, 20, 30, 40, 50\}$, we generated a set of \textbf{100 random DAG input files} using a probabilistic graph generator that maintains acyclic structure and realistic edge densities. Each DAG represents a periodic job with a fixed period $D$, random task computation times, and varying inter-task communication demands. The goal is to simulate a broad and diverse workload profile that mirrors real-world variations in edge deployments. Three key parameters were varied to analyze the proposed scheduler:
\begin{enumerate}
    \item \textbf{Edge Server Occupancy:} the percentage of MEC resource already occupied by background workloads.
    \item \textbf{Communication-to-Computation Ratio (CCR):} defines the relative cost of inter-task communication versus task execution.
    \item \textbf{Number of Available Processors:} the computing resources available at the edge for executing DAG tasks.
\end{enumerate}

\subsection{Impact of Edge Server Occupancy}

\begin{figure}[h]
    \centering
    \includegraphics[width=0.45\textwidth]{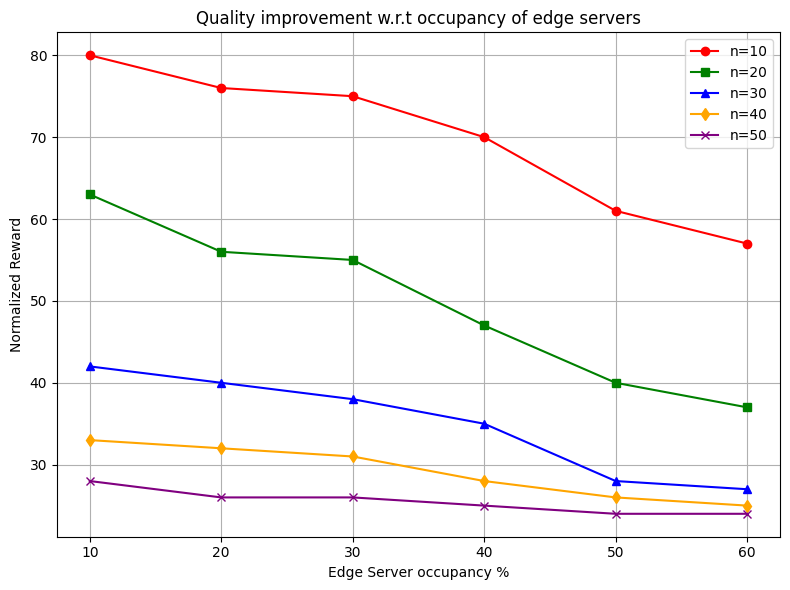}
    \caption{Normalized Reward vs. Edge Server Occupancy}
    \label{fig:occupancy}
\end{figure}

Figure~\ref{fig:occupancy} presents the variation of normalized reward with respect to edge server occupancy, which ranges from 10\% to 60\%. Higher occupancy reduces the residual capacity available for new DAGs.

\begin{itemize}
    \item At low occupancy (10\%), small DAGs ($n=10$) achieve normalized rewards close to 80\%, thanks to better task placement and minimal contention.
    \item As occupancy increases, the normalized reward declines for all DAG sizes due to reduced flexibility in task-to-VM mapping and increased competition for resources.
    \item DAGs with larger $n$ (e.g., $n=50$) consistently achieve lower NR (25–30\%), primarily because more tasks are squeezed into limited residual capacity, increasing the chance of deadline violations.
    \item However, larger DAGs show relatively stable performance with respect to occupancy variations, indicating robustness at scale.
\end{itemize}

This reveals a trade-off between scale and sensitivity: while smaller DAGs yield higher QoS under light load, they are more fragile under high contention. Larger DAGs, though limited in maximum reward, offer more graceful degradation.

\subsection{Impact of Communication-to-Computation Ratio (CCR)}

\begin{figure}[h]
    \centering
    \includegraphics[width=0.45\textwidth]{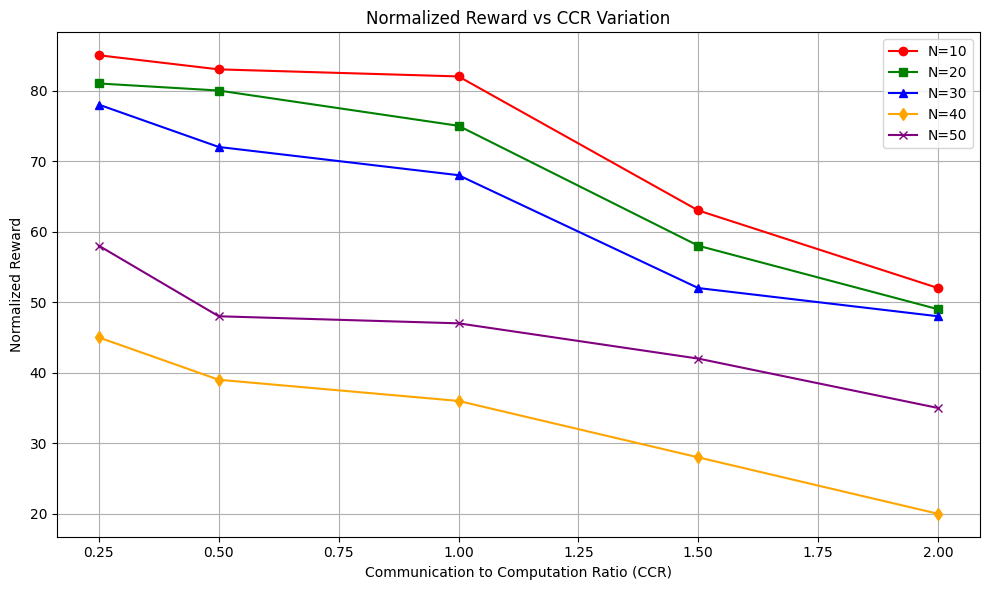}
    \caption{Normalized Reward vs. Communication-to-Computation Ratio (CCR)}
    \label{fig:ccr}
\end{figure}

Figure~\ref{fig:ccr} analyzes the effect of increasing CCR on normalized reward. Higher CCR implies greater cost for task communication (e.g., data exchange, synchronization), making scheduling more challenging.

\begin{itemize}
    \item For low CCR (0.25), DAGs benefit from low communication penalties, with small DAGs ($n=10$, $n=20$) achieving NR $\approx$ 85–80\%.
    \item As CCR increases to 1.5 and 2.0, the NR degrades significantly across all DAG sizes. This is due to longer communication delays causing downstream task delays or deadline violations.
    \item The penalty is more prominent in large DAGs, where inter-task dependencies are higher.
    \item The best-performing scenarios are those with low CCR and smaller DAGs, where more service levels can be fulfilled despite communication overhead.
\end{itemize}

These results underline the importance of communication-aware scheduling. DAGs with tight coupling and high CCR require more sophisticated placement strategies, such as co-locating dependent tasks to minimize delay.

\subsection{Effect of Number of Processors}

\begin{figure}[h]
    \centering
    \includegraphics[width=0.45\textwidth]{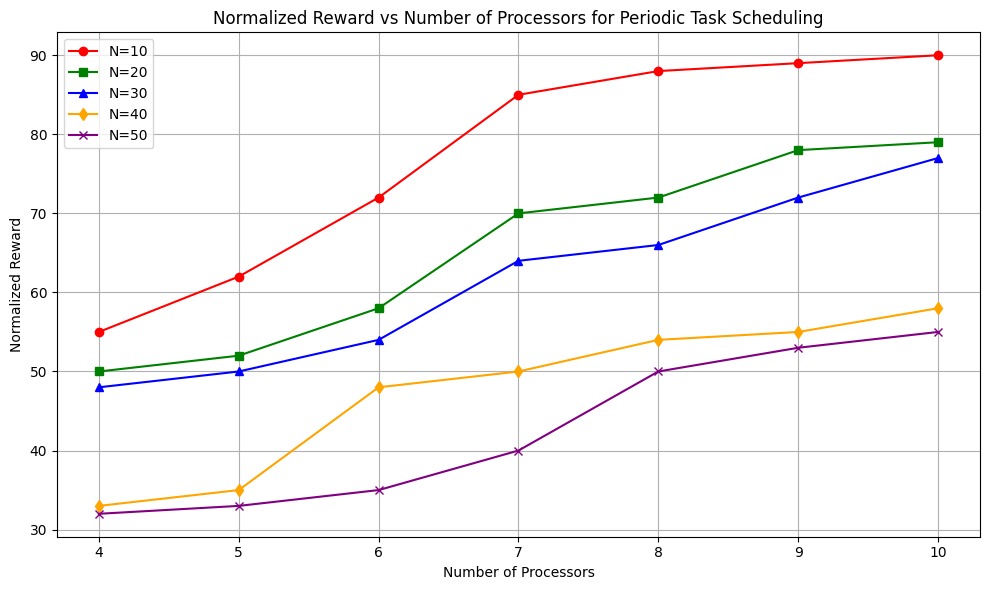}
    \caption{Normalized Reward vs. Number of Processors}
    \label{fig:processors}
\end{figure}

Figure~\ref{fig:processors} investigates how increasing the number of processors impacts normalized reward.

\begin{itemize}
    \item The reward increases monotonically with the number of processors, affirming that higher parallelism enables better scheduling flexibility.
    \item Smaller DAGs ($n=10$, $n=20$) benefit substantially, achieving NR as high as 90\% when 10 processors are available. These DAGs can be more easily parallelized.
    \item Larger DAGs ($n=40$, $n=50$) also show reward gains, though to a lesser extent, due to inherent serial dependencies and communication costs.
    \item The reward increase plateaus beyond 8 processors for all DAGs, suggesting diminishing returns from adding more processors.
\end{itemize}

These findings highlight the necessity of sufficient compute provisioning in latency-critical MEC deployments. Even under constrained resource pools, scaling processors significantly improves task feasibility and overall QoS.

\subsection{Summary of Insights}

\begin{itemize}
 \item \textbf{Randomized Sampling:} Each result is averaged over 100 randomized DAGs per size, ensuring robustness and generalizability of conclusions.
    \item \textbf{Edge Server Occupancy:} Strong inverse correlation with reward. Small DAGs yield higher QoS under low load but degrade faster.
    \item \textbf{Edge server occupancy} has a strong inverse correlation with reward; lower occupancy allows high-quality schedules, especially for small DAGs.
    \item \textbf{CCR variation} penalizes reward as communication becomes a bottleneck. This requires task co-location strategies and communication-aware version selection.
    \item \textbf{Processor scaling} is highly effective for improving reward, though benefit saturates beyond a certain threshold due to non-parallelizable dependencies.
\end{itemize}

Overall, the proposed scheduler adapts effectively to changing MEC conditions and task graph sizes. By modeling scheduling as a reward-maximizing problem constrained by periodicity, precedence, and resource usage, it provides a robust and scalable solution for emerging real-time edge applications.






    


\section{Conclusion and Future Works}
In conclusion, this research introduces a novel, QoS-aware task scheduling framework tailored for periodic Directed Acyclic Graphs (DAGs) in heterogeneous, pre-occupied Mobile Edge Computing (MEC) environments. The proposed solution builds upon the Heterogeneous Earliest Finish Time (HEFT) algorithm by introducing modifications to handle pre-allocated workloads and optimize scheduling within available idle gaps. By leveraging task versioning and dynamic gap-aware mapping, the framework achieves high processor utilization while maximizing aggregate QoS — without violating constraints of periodicity, resource availability, or task precedence.

The Periodic Scheduler component harmonizes schedules from multiple DAGs by computing a common hyperperiod, ensuring that execution is deterministic and conflict-free across recurring cycles. This makes the approach particularly valuable in latency-sensitive, real-time systems.

\subsection*{Key Takeaways}
\begin{itemize}
    \item Efficiently integrates new periodic DAGs alongside pre-existing workloads on shared processors.
    \item Enhances task quality where possible by rescheduling within residual time windows.
    \item Provides a scalable and modular solution suitable for both small and large DAGs.
    \item Demonstrates robustness under varying network occupancy, communication cost ratios (CCR), and processor availability.
\end{itemize}

\subsection*{Real-World Applications}
\begin{itemize}
    \item \textbf{Smart Traffic Monitoring:} Real-time integration of emergency traffic updates (e.g., accidents or road closures) alongside scheduled traffic flow analytics.
    \item \textbf{Healthcare Monitoring:} Periodic vitals monitoring with added emergency ECG analysis during suspected anomalies, running concurrently.
    \item \textbf{Smart Surveillance:} Scheduled camera feeds running concurrently with high-quality forensic enhancement jobs triggered by anomaly detection.
    \item \textbf{Autonomous Drones:} Periodic sensor and telemetry data collection with dynamic flight re-routing tasks during emergencies or signal loss.
\end{itemize}

\subsection*{Future Work}
\begin{itemize}
    \item \textbf{Energy-Aware Scheduling:} Extending the current algorithm to factor in energy profiles and thermal limits, enabling scheduling decisions that reduce energy consumption or carbon footprint.
    \item \textbf{Fault-Tolerant Scheduling:} Incorporating failure detection and rollback mechanisms to support high-availability scheduling in safety-critical environments.
    \item \textbf{Online Scheduling Extensions:} Real-time adaptations of the algorithm that respond to dynamically arriving DAGs or updated processor availability.
\end{itemize}

Overall, this work lays a robust and extensible foundation for quality-driven, resource-aware, and real-time task scheduling in edge computing ecosystems. It bridges theoretical scheduling optimization with practical system deployment requirements, making it a strong candidate for real-world adoption in smart cities, healthcare, robotics, and Industry 4.0 applications.

\vspace{12pt}

\end{document}